\documentclass[12pt,preprint]{aastex}
\usepackage{graphics}
\usepackage{bm}
\usepackage{epsfig}
\usepackage{graphicx}
\usepackage{subfigure}

\begin{document}

\title{A General Relativistic Magnetohydrodynamic Model of High Frequency
Quasi-periodic Oscillations in Black Hole Low-Mass X-ray Binaries}

\author{Chang-Sheng Shi\altaffilmark{1,2,3} and Xiang-Dong
Li\altaffilmark{1,2}}

\altaffiltext{1}{Department of Astronomy, Nanjing University,
Nanjing 210093, China; E-mail: scs1217@gmail.com, lixd@nju.edu.cn}
\altaffiltext{2}{Key Laboratory of Modern Astronomy and Astrophysics
(Nanjing University), Ministry of Education, Nanjing 210093, China}
\altaffiltext{3}{College of Material Science and Chemical
Engineering, Hainan University, Hainan 570228, China}

\begin{abstract}
We suggest a possible explanation for the high frequency
quasi-periodic oscillations (QPOs) in black hole low mass X-ray
binaries. By solving the perturbation general relativistic
magnetohydrodynamic equations, we find two stable modes of the
Alf\'ven wave in the the accretion disks with toroidal magnetic
fields. We suggest that these two modes may lead to the double high
frequency QPOs if they are produced in the transition region between
the inner advection dominated accretion flow and the outer thin
disk. This model  naturally accounts for the $3:2$ relation for the
upper and lower frequencies of the QPOs, and the relation between
the black hole mass and QPO frequency.

\end{abstract}


\section{Introduction}

Low-mass X-ray binaries (LMXBs) are binary systems consisting of a
neutron star (NS) or black hole (BH) accreting from a low-mass
($\lesssim 1 M_{\sun}$) companion star. X-ray emission of LMXBs
often shows fast X-ray variability in the form of high frequency
quasi-periodic oscillations (HFQPOs), which frequently appear in
pairs in certain state simultaneously (van der Klis 2006).
Abramowicz \& Klu\'zniak (2001) pointed out that the frequency ratio
of the twin-peak HFQPOs in the BH source GRO J1655-40 equals 3/2,
and that this commensurability of frequencies may be a signature of
a non-linear resonance. Later, Abramowicz et al. (2003) found a
signature of the same commensurable ratio in the twin-peak HFQPOs
observed in an NS-LMXB, Sco X-1. Based on this observational
evidence, Klu{\'z}niak and Abramowicz argued in several papers
(Klu{\'z}niak \& Abramowicz 2001; Klu{\'z}niak, Abramowicz, \& Lee
2004; Klu{\'z}niak et al. 2004) that the twin-peak HFQPOs in both BH
and NS sources are due to the same physical mechanism --- a
non-linear parametric resonance in accretion disk global
oscillations.

However, while for the BH sources the commensurable ratio 3/2 was
quickly confirmed and generally accepted (e.g. Remillard \&
McClintock 2006), the presence of the same commensurability in the
NS sources is denied by several experts (e.g. Boutelier et al.,
2009). There is no consensus whether the nature of the twin-peak
HFQPOs in the two types of LMXBs is the same. We have proposed a
mechanism for the twin kilohertz QPOs in NS-LMXBs using the
magnetohydrodynamic (MHD) Alf\'ven wave oscillations, and the
results seem to fit the observation well (Li \& Zhang 2005; Shi \&
Li 2009). In this paper we focus on an MHD explanation of the HFQPOs
in BH-LMXBs.

Barret et al. (2005) measured the quality factor Q for the HFQPOs
measured in the NS-LMXB 4U 1608-52, and found that Q $\thicksim$
200. They argued that such high coherency is impossible to achieve
from kinematic effects in orbital motion of hot spots, clumps or
other similar features located at the accretion disk surface,
because these features are too quickly sheared out by the
differential rotation. Although orbital motion cannot explain the
the HFQPOs in LMXBs, the frequencies of several fluid oscillatory
modes are expressed by the three characteristic orbital frequencies:
the ``Keplerian" frequency, the ``radial" epicyclic frequency, and
the ``vertical" epicyclic frequency. In the Kerr metric, these three
orbital frequencies and the Lense-Thirring ``frame-dragging"
frequency have been listed (e.g. Perez et al. 1997). Several HFQPOs
models use their ratios (in various combinations) to explain the
observed 3/2 commensurability. Cui, Zhang, \& Chen (1998) suggested
the Lense-Thirring nodal precession frequency near the inner stable
circular orbit (ISCO) radius as the lower HFQPO frequency, such as
the 300 Hz QPOs in GRO J1655-40. The relativistic precession model
of Stella et al. (1999) applies to both BH and NS sources; the
pariastron precession frequency and the Keplerian frequency were
taken as the lower and upper frequencies of the twin HFQPOs,
respectively, whereas the QPOs at much lower frequencies were
interpreted in terms of the Lense-Thirring nodal precession
ferquency. Wang et al. (2003, 2005) suggested that a
non-axisymmetric magnetic coupling of a rotating BH with its
surrounding accretion disk coexists with the Blandford-Znajek
process. The two frequencies were supposed as the Keplerian
frequencies of two hotspots, one near the inner edge of the disk and
the other somewhere outside, respectively.

Wagoner et al. (2001) considered the modes of the diskoseismic wave,
such as g-modes, p-modes and c-modes as the explanation of the
HFQPOs. They estimated the masses and angular momenta of some BHs
with the measured frequencies of the HFQPOs when the g-modes or
c-modes were selected. Rezzolla et al. (2003) discussed the
inertial-acoustic modes in a small-size torus very close to the
horizon of the BH while the centrifugal and pressure gradients were
selected as the only restoring forces. In this model the black hole
spin had to be very close to the maximal value to produce the $3:2$
ratio. Tassive \& Bertschinger (2008) investigated the kinematic
density waves in the accretion disks when nothing but the gravity
was considered as the restoring force, and several discrete radii
were adopted. Several modes in pairs close the ratio ($3:2$) could
be got, but the correct frequencies couldn't be reproduced. Similar
to the parametric resonance models, it is difficult to explain why
other modes such as the fundamental frequencies weren't observed
except the two modes in pairs.

In modeling HFQPOs in BH-LMXBs two points need to be mentioned.
First, in most models the $3:2$ ratio was often overemphasized and
substituted into those models directly. In fact, the  $3:2$ ratio of
the twin HFQPOs in BH-LMXBs isn't rigorous but approximate. Second,
the HFQPO frequencies were often considered invariable so that these
frequencies (168, 113 Hz and 67, 41Hz in GRS 1915+105, Remillard
(2004); 450, 300 Hz in GRO J1655-40, Strohmayer (2001a); 276, 184 Hz
in XTE J1550-564, Miller et al. (2001); 240, 165 Hz in H1743-322,
Homan et al. (2005)) were used to estimate the parameters of the BHs
such as the masses and the spins. In fact, the frequencies are
stable, i.e. there are small varitions in them, rather than
invariable (Miller et al. 2001; Remillard et al. 1999; Morgan et al.
1997; Strohmayer 2001b).

This paper is organized as follows. In Section 2 we suggest the
basic model and get two stable modes of the Alf\'ven wave by solving
the general relativistic magnetohydrodynamic (GRMHD) equations of
the perturbed plasma in BH accretion disks. In Section 3 we compare
the results with the observations and discuss their possible
implications.

\section{The GRMHD model of the HFQPOs in BH-LMXBs}

We consider that the HFQPOs of BH-LMXBs result from GRMHD waves
caused by perturbations in the disk. According to \S2.1 we can find
that only one type of GRMHD wave, the steady Alf\'ven wave, can
spread along the magnetic field lines to the energy release region.
The simulation by Koide (2002) also shows that a torsional Alf\'ven
wave can be generated by the rotational dragging of space. We assume
that the puny disturbance doesn't change the metric of the
space-time.

\subsection{ The modes of the GRMHD wave}

The oscillation modes of plasma in accretion disks with or without
magnetic field, such as g-modes, p-modes, and c-modes, have ever
been investigated (Wagoner et al. 2001; Fu \& Lai 2009; Lai 2009).
Here we discuss the GRMHD wave modes in ideal adiabatic
magnetofluid. In the fiducial observer (FIDO) frame, which is a
locally inertial frame, the line element can be written as $ ({\rm
d}s)^2 =- (c{\rm d}t)^2 + \sum\limits_{i = 1}^3 { ({\rm d}x^i)^2} $,
where $c$ is the speed of light in vacuum, the Roman indices ($i$)
run from 1 to 3, and $(ct, x^1, x^2, x^3)$ are the coordinates of
the FIDO frame. The metric of the Kerr space-time in the
Boyer-Lindquist coordinates ($ct', r, \theta, \varphi$) is
introduced and the line element for the observer at infinity (i.e.
in the laboratory frame) can be written as,
\begin{equation}
\label{eq1} ({\rm d}s)^2 =- h_0^2 (c{\rm d}t')^2 + \sum\limits_{i =
1}^3 {[h_i^2 ({\rm d}x^{i}{'})^2 - 2h_i^2 \omega_i {\rm d}t'{\rm
d}x^{i}{'}]},
\end{equation}
where
\begin{equation}
\label{eq2} h_0^2 = 1-\frac{2rr_{g}}{\Sigma},\ h_1^2 =
\frac{\Sigma}{\Delta},\ h_2^2 = \Sigma, h_3^2 =
\frac{A}{\Sigma}\sin^2 \theta,\ \omega_1=\omega_2=0, \
\omega_3=\frac{2carr_{g}^2}{A}.
\end{equation}
Here $(ct', x^1{'}, x^2{'}, x^3{'})$ are the coordinates of the
laboratory frame,  $h_0$, $h_i$ and $\omega_i$ are the metrics of
the Kerr space-time,  $r_g= GM/c^2$, $a=Jc/GM^2$ ($M$ and $J$ are
the mass and the angular momentum of the BH respectively), $G$ is
gravitational constant, $\Sigma=r^2+a^2 r_g^2 \cos^2 \theta$,
$\Delta=r^2-2rr_g+a^2 r_g^2$, and
$A=(r^2+a^2r_g^2)^2-a^2r_g^2\sin^2\theta\Delta$. Since our
discussion is limited to the accretion disk we take $\theta=\pi/2$
and get $\Sigma=r^2$ and $A=(r^2+a^2r_g^2)^2-a^2r_g^2\Delta$, where
$r$ is the distance of the plasma from the BH. Besides that, the
lapse function and the shift velocity can be expressed as,
\begin{equation}
\label{eq3} \alpha = \sqrt {h_0 ^2 + \sum\limits_{i = 1}^3
{(\frac{h_i \omega _i }{c})^2} }
=\sqrt{\frac{r^4-2r^3r_g+a^2r^2r_g^2}{r^4+a^2r^2r_g^2+2a^2rr_g^3}},
\end{equation}
and
\begin{equation}
\label{eq4}\beta^i=\frac{h_i\omega_i}{c\alpha} \mbox{ or }
\bm{\beta}=(\beta_1, \beta_2,
\beta_3)=(0,0,\frac{2ar_g^2}{r\sqrt{r^2-2rr_g+a^2r_g^2}}),
\end{equation}
where $\bm{\beta}$ is a vector parallel to the toroidal velocity of
the plasma.

We begin with the form of 3+1 split of the GRMHD equations as in
Koide (2003). Previous investigations (Ruzmaikin et al. 1979; Tout
et al. 1992; Ruediger et al. 1995; Hawley et al. 2000; Moss et al.
2004; Hirose et al. 2004) have shown that in the accretion disk
around a BH the toroidal component of the magnetic field may be much
stronger than the poloidal component, i.e., $B_r \ll B_\varphi$ and
$B_\theta \ll B_\varphi$. In our analysis we also assume $v_r \ll
v_\varphi$ and $v_\theta \ll v_\varphi$. The equations in the FIDO
frame are then written as follows:
\begin{equation}
\label{eq5} \frac{\partial (\gamma\rho)}{\partial t} = - \nabla
\cdot [\alpha \gamma\rho({\rm {\bf v}} + c{{\rm {\bm \beta}}})],
\end{equation}
\begin{equation}
\label{eq6} \frac{\partial {\rm {\bf P}}}{\partial t} = - \nabla
\cdot [\alpha ({\rm \widetilde{\bf T}} + c{\rm {\bf \bm{\beta} P}})]
- (\varepsilon + \gamma\rho c^2)\nabla \alpha + \alpha {\rm {\bf
f}_{curv}} - {\rm {\bf P}} \cdot {\rm \widetilde{\bf \sigma }},
\end{equation}
\begin{equation}
\label{eq7} \frac{\partial \varepsilon }{\partial t} = - \nabla
\cdot [\alpha (c^2{\rm {\bf P}} - \gamma\rho c^2{\rm {\bf v}} +
\varepsilon c{\rm {\bf \bm{\beta} }})] - (\nabla \alpha ) \cdot
c^2{\rm {\bf P}} - {\rm \widetilde{\bf T}}:{\rm \widetilde{\bf
\sigma }},
\end{equation}
\begin{equation}
\label{eq8} \frac{\partial {\rm {\bf B}}}{\partial t} = - \nabla
\times [\alpha ({\rm {\bf E}} - c{\rm {\bf \bm{ \beta} }}\times {\rm
{\bf B}})],
\end{equation}
\begin{equation}
\label{eq9} \nabla \cdot {\rm {\bf B}} = 0,
\end{equation}
\begin{equation}
\label{eq10} {\rm {\bf E}} + {\rm {\bf v}}\times {\rm {\bf B}} = 0,
\end{equation}
\begin{equation}
\label{eq11}p\rho^{-\Gamma} = {\rm constant},
\end{equation}
where  {\bf v} is the velocity of the plasma, $\rho$ the plasma
density, $p$ the barometric pressure, $\Gamma$ the adiabatic index,
$\gamma$ the Lorentz factor, $\bf E={{\bf E}^{'}}/{\sqrt{\mu_0}}$,
$\bf B={{\bf B}^{'}}/{\sqrt{\mu_0}}$ (here ${\bf E}^{'}$ is the
electric field, ${\bf B}^{'}$ the magnetic field, and $\mu_0$ the
magnetic permeability in the vacuum), respectively. The bold
characters denote vectors, and the superscript $\sim$ corresponds to
tensors. The energy-momentum tensor is,
\begin{equation}
\label{eq12} {\rm \widetilde{\bf T}}=(p+ \frac{B^2}{2}+
\frac{E^2}{2c^2})\rm\widetilde{\bf I}+ \frac{\psi}{\emph{c}^2}
\gamma^2{\bf V} {\bf V}-{\bf B}{\bf B} -
 \frac{1}{\emph{c}^2}{\bf E} {\bf E},
\end{equation}
where $\psi=\rho \emph{c}^2+\frac{
 \Gamma \emph{p}}{\Gamma-1}$ is the relativistic enthalpy density. The
equivalent momentum density and energy density are,
\begin{equation}
\label{eq13} {\rm {\bf P}}= \frac{\psi}{c^2}\gamma^2{\bf V} +
\frac{1}{c^2}{{\bf E} \times{ \bf B }}.
\end{equation}
and
\begin{equation}
\label{eq14} \varepsilon=\psi\gamma^2- p -\gamma \rho c^2+
 \frac{B^2}{2}+ \frac{E^2}{2c^2},
\end{equation}
respectively. Equations (5), (6), (7), (8), and (9) correspond to
the continuity equation, the momentum conservation equation, the
energy conservation equation, the law of electromagnetic induction,
and the equation of no divergence, respectively. Equations (10) and
(11) are the equations of state for the infinite electrical
conductivity and the adiabatic condition.  The other physical
quantities are,
\begin{equation}
\label{eq15} {\bf f}_{\rm curv} \equiv \sum\nolimits_j {(G_{ij}
T^{ij} - G_{ji} T^{jj})},\   \sigma _{ij} \equiv (h_i / h_j
)(\partial \omega _i / \partial x^j),\  G_{ij} \equiv - (1 / h_i h_j
)\times (\partial h_i /\partial x^j).
\end{equation}
We can simplify the above two tensors in the accretion disk in the
Kerr space-time as,
\begin{equation}
\label{eq16} \sigma _{ij} =�\left\{ {{\begin{array}{*{20}c}
 0 \hfill & 0 \hfill & 0 \hfill \\
 0 \hfill & 0 \hfill & 0 \hfill \\
 { - \frac{2acr_g^2 (3r^2 + a^2r_g^2 )\sqrt {r^2 + a^2r_g^2 +
\textstyle{{2a^2r_g^3 } \over r}} }{\sqrt {\textstyle{{r^2} \over
{r^2 + a^2r_g^2 - 2rr_g }}} (r^3 + ra^2r_g^2 + 2a^2r_g^3 )^2}}
\hfill & 0 \hfill &
0 \hfill \\
\end{array} }} \right\} = \left( {{\begin{array}{*{20}c}
 0 \hfill \\
 0 \hfill \\
 1 \hfill \\
\end{array} }} \right)({\begin{array}{*{20}c}
 {\frac{h_3 }{h_1 }\frac{\partial \omega _3 }{\partial x^1}} \hfill & 0
\hfill & 0 \hfill \\
\end{array} }),
\end{equation}
and
\begin{equation}
\label{eq17} G_{ij} = - \left\{ {{\begin{array}{*{20}c}
 {\frac{r_g (a^2r_g - r)}{r^2\sqrt {r^2 - 2rr_g + a^2r_g^2 } }} \hfill & 0
\hfill & 0 \hfill \\
 {\frac{\sqrt {r^2 - 2rr_g + a^2r_g^2 } }{r^2}} \hfill & 0 \hfill & 0 \hfill
\\
 {\frac{(r^3 - a^2r_g^3 )\sqrt {r^2 - 2rr_g + a^2r_g^2 } }{(r^3 + ra^2r_g^2
+ 2a^2r_g^3 )r^2}} \hfill & 0 \hfill & 0 \hfill \\
\end{array} }} \right\},
\end{equation}
where we define the vector ${\bf N} \equiv [(h_3 /h_1) (\partial
\omega _3 /\partial x^1),0,0]$. The magnetized accretion torus with
a toroidal magnetic field around a Kerr black hole can exist stably
(Komissarov 2006), and the GRMHD equations in the steady state can
be expressed as,
\begin{equation}
\label{eq18}\frac{\partial (\gamma \rho _0 )}{\partial t} = - \nabla
\cdot [\alpha \gamma \rho _0 ({\rm {\bf v} _0} + c{\rm {\bm \beta
}})],
\end{equation}
\begin{equation}
\label{eq19}\frac{\partial {\rm {\bf P}_0}}{\partial t} = - \nabla
\cdot [\alpha ({\rm \widetilde{\bf T}}_0 + c{\rm {\bf \bm{\beta}
P}_0})] - (\varepsilon_0 + \gamma \rho_0 c^2)\nabla \alpha + \alpha
{\rm {\bf f}}_{\rm curv,0} - {\rm {\bf P}_0} \cdot {\rm
\widetilde{\bf \sigma }},
\end{equation}
\begin{equation}
\label{eq20  }\frac{\partial \varepsilon_0 }{\partial t} = - \nabla
\cdot [\alpha (c^2{\rm {\bf P}_0} - \gamma \rho_0 c^2{\rm {\bf v}_0}
+ \varepsilon_0 c{\rm {\bm \beta }})] - (\nabla \alpha ) \cdot
c^2{\rm {\bf P}_0} - {\rm \widetilde{{\bf T}}_0}:{\rm
\widetilde{{\bf \sigma }}},
\end{equation}
\begin{equation}
\label{eq21  }\frac{\partial {\rm {\bf B}_0}}{\partial t} = - \nabla
\times [\alpha ({\rm {\bf E}_0} - c{\rm {\bm \beta }}\times {\rm
{\bf B}_0})],
\end{equation}
\begin{equation}
\label{eq22  }\nabla \cdot {\rm {\bf B}_0} = 0,
\end{equation}
\begin{equation}
\label{eq23  }{\rm {\bf E}_0} + {\rm {\bf v}_0}\times {\rm {\bf
B}_0} = 0,
\end{equation}
\begin{equation}
\label{eq24  }p_0\rho_0^{-\Gamma} = {\rm constant},
\end{equation}
where the subscript 0 denotes the variables in steady state. Next we
consider the GRMHD equations after the plasma is perturbed slightly,
\begin{equation}
\label{eq25} \frac{\partial (\gamma \hat{\rho})}{\partial t} = -
\nabla \cdot [\alpha \gamma \hat{\rho}({\rm \hat{{\bf v}}} + c{{\rm
{\bm \beta}}})],
\end{equation}
\begin{equation}
\label{eq26} \frac{\partial {\rm \hat{{\bf P}}}}{\partial t} = -
\nabla \cdot [\alpha ({\rm \hat{\widetilde{\bf T}}} + c{\rm {\bf
\bm{\beta} \hat{P}}})] - (\hat{\varepsilon} + \gamma\hat{\rho}
c^2)\nabla \alpha + \alpha {\rm \hat{{\bf f}}}_{\rm curv} -
{\rm\hat{ {\bf P}}} \cdot {\rm \widetilde{\bf \sigma }},
\end{equation}
\begin{equation}
\label{eq27} \frac{\partial \hat{\varepsilon} }{\partial t} = -
\nabla \cdot [\alpha (c^2{\rm \hat{{\bf P}}} - \gamma\hat{\rho}
c^2{\rm \hat{{\bf v}}} + \hat{\varepsilon} c{\rm {\bf \bm{\beta}
}})] - (\nabla \alpha ) \cdot c^2{\rm \hat{\bf P}} - {\rm
\hat{\widetilde{\bf T}}}:{\rm \widetilde{\bf \sigma }},
\end{equation}
\begin{equation}
\label{eq28} \frac{\partial {\rm \hat{{\bf B}}}}{\partial t} = -
\nabla \times [\alpha ({\rm\hat{ {\bf E}}} - c{\rm {\bf \bm{ \beta}
}}\times {\rm \hat{{\bf B}}})],
\end{equation}
\begin{equation}
\label{eq29} \nabla \cdot {\rm\hat{{\bf B}}} = 0,
\end{equation}
\begin{equation}
\label{eq30} {\rm \hat{{\bf E}}} + {\rm\hat{ {\bf v}}}\times {\rm
\hat{{\bf B}}} = 0,
\end{equation}
\begin{equation}
\label{eq31}\hat{p}\hat{\rho}^{-\Gamma} = p_0\rho_0^{-\Gamma},
\end{equation}
where $\hat{\bm v} = {{\bm v}}{ }_0 + { {\bm v}}_{s} $, $\hat{\rho}
= \rho_0 + \rho_{s}$, $\hat{p} = p_0 + p_{ s}$, $\hat{\varepsilon} =
\varepsilon_0 + \varepsilon_{s}$, $\hat{\bm E} = { {\bm E}}{ }_0 +
{{\bm E}}_{ s}$, $\hat{\bm B} = { {\bm B}}_{ {\bm 0}} + {{\bm B}}{
}_{{s}}$, $\hat{\bm P} = {{\bm P}}_0 + {{\bm P}}_{s}$, $\hat{\bm
f}_{\rm curv} = {{\bm f}}_{\rm curv,0 } + {{\bm f}}_{{\rm curv},s}$,
and $\hat{\widetilde{\bm T}} = {\widetilde{\bm T}}_0 +
{\widetilde{\bm T}}_{s}$, with the subscript $s$ denoting the
perturbed quantities ($v_{ s} \ll {v_0 }$, $E_{ s}\ll E_0$, $B_{
s}\ll B{ }_0$, $\rho_{ s}\ll \rho_0$, and $p_{s}\ll p_0$) and with
the superscript $\hat{}$ denoting the variables after the
disturbance. By combining Eqs.~ (18)-(31) we get the equations about
the perturbed quantities in the first-order approximation,
\begin{equation}
\label{eq32}\frac{\partial (\gamma \rho _s )}{\partial t} = - \nabla
\cdot [\alpha \gamma (\rho _0 {\rm {\bf v}}_{{s}} + \rho _s
{\rm {\bf v}} + \rho _s c{\rm {\bf \beta }})],
\end{equation}
\begin{equation}
\label{eq33}\frac{\partial {\rm {\bf P}}_{{s}} }{\partial t}
= - \nabla \cdot [\alpha ({\widetilde{\rm {\bf T}}}{ }_{ { s}} + c{\rm
{\bf \beta P}}_{ { s}} )] - (\varepsilon _s + \gamma \rho _s
c^2)\nabla \alpha + \alpha {{ f}}_{{\rm curv},s} - {\rm {\bf
P}}_{ {s}} \cdot {\rm {\bf \sigma }},
\end{equation}
\begin{equation}
\label{eq34}\frac{\partial \varepsilon _s }{\partial t} = - \nabla
\cdot [\alpha (c^2{\rm {\bf P}}_{{ s}} - \gamma c^2\rho _0
{\rm {\bf v}}_{{ s}} - \gamma c^2\rho _s {\rm {\bf v}}_0 +
\varepsilon _s c{\rm {\bf \beta }})] - (\nabla \alpha ) \cdot
c^2{\rm {\bf P}}_{{ s}} - {\rm {\bf T}}_{ {s}} :{\rm
{\bf \sigma }},
\end{equation}
\begin{equation}
\label{eq35}\frac{\partial {\rm {\bf B}}_{ { s}} }{\partial t}
= - \nabla \times [\alpha ({\rm {\bf E}}_{{ s}} - c{\rm {\bf
\beta }}\times {\rm {\bf B}}_{{ s}} )],
\end{equation}
\begin{equation}
\label{eq36}\nabla \cdot {\rm {\bf B}}_{ { s}} = 0,
\end{equation}
\begin{equation}
\label{eq37}{\rm {\bf E}}_{ { s}} + {\rm {\bf v}}_0 \times
{\rm {\bf B}}_{{ s}} + {\rm {\bf v}}_{ { s}} \times
{\rm {\bf B}}_0 = 0,
\end{equation}
\begin{equation}
\label{eq38} p_s = \frac{\Gamma p_0 }{\rho _0 }\rho _s.
\end{equation}
With Eqs.~(12)-(14) we can get the perturbed energy-momentum tensor
(${\widetilde{\bm T}}{ }_s$), equivalent momentum density (${\bm
P}_s$) and equivalent energy density (${\bm \varepsilon}_s$) in the
same way, by defining $\psi_0 =\rho_0 \emph{c}^2+\frac{
 \Gamma \emph{p}_0}{\Gamma-1}$ and $\psi_s=\rho_s \emph{c}^2+\frac{
 \Gamma \emph{p}_s}{\Gamma-1}$,
\begin{eqnarray}
\label{eq39} {\rm \widetilde{\bf T}}_s =(p_s + {\bf B}_0 \cdot {\bf
B}_s+ \frac{1}{c^2}{\bf E}_0 \cdot {\bf E}_s)\rm\widetilde{\bf I}+
\frac{\psi_s}{\emph{c}^2}\gamma^2{\bf V}_0 {\bf V}_0 \nonumber +
\frac{\psi_0}{\emph{c}^2}\gamma^2({\bf V}_0 {\bf V}_s+{\bf V}_s {\bf
V}_0)\\
 -({\bf B}_0{\bf B}_s+{\bf B}_s{\bf B}_0 )-
 \frac{1}{\emph{c}^2}({\bf E}_0 {\bf E}_s+{\bf E}_s {\bf E}_0),
\end{eqnarray}
\begin{equation}
\label{eq40} {\rm {\bf P}_s}= \frac{\gamma^2}{c^2}(\psi_0 {\bf v}_s
+ \psi_s {\bf v}_0) + \frac{1}{c^2}{{\bf E}_s \times{ \bf B }_0} +
\frac{1}{c^2}{{\bf E}_0 \times{ \bf B }_s},
\end{equation}
\begin{equation}
\label{eq41} \varepsilon=\psi_s \gamma^2- p_s -\gamma \rho_s c^2+
{\bf B}_0 \cdot {\bf B}_s + \frac{{\bf E}_0 \cdot {\bf E}_s}{c^2}.
\end{equation}
After carrying out Fourier transformation ($e^{i{{\bf k}\cdot{\bf r}}-i\omega t}$)
for Eqs.~(32)-(36) and substituting Eqs.~(37) and (38) into them, we get the
following equations when ${\bf v}_0 {\parallel} {\bf B}_0 {
\parallel}{\bm \beta}$ is considered,
\begin{equation}
\label{eq42} [\omega   - \alpha  ({\rm {\bf k}} \cdot {\rm {\bf
v}}_0 ) - \alpha c ({\rm {\bf k}} \cdot {\rm {\bf {\bm \beta}
}})]\rho _{ s} = \alpha  \rho _0 ({\rm {\bf k}} \cdot {\rm {\bf
v}}_{ { s}} ),
\end{equation}
\begin{equation}
\label{eq43} \omega {\rm {\bf P}}_{ { s}} - \alpha c({\rm {\bf
k}} \cdot {\rm {\bf {\bm \beta }}}){\rm {\bf P}}_{ { s}} =
\alpha {\rm {\bf k}} \cdot {\rm \widetilde {\bf T}}{ }_{ { s}}
+ \alpha (\varepsilon_{ s} + \gamma \rho _{ s} c^2){\rm {\bf k}} +
i\alpha {\rm {\bf f}}_{{\rm curv},s} - i{\rm {\bf P}}_{ { s}} \cdot
{\rm \widetilde {\bf \sigma }},
\end{equation}
\begin{equation}
\label{eq44} \omega \varepsilon_{ s} - \alpha c({\rm {\bf k}} \cdot
{\rm {\bf \bm{\beta} }})\varepsilon_{ s} = 2\alpha c^2({\rm {\bf k}}
\cdot {\rm {\bf P}}_{ s} ) - \alpha \gamma c^2\rho _0 ({\rm {\bf k}}
\cdot {\rm {\bf v}}_{ s} ) - \alpha \gamma c^2({\rm {\bf k}} \cdot {\rm
{\bf v}}_0 )\rho _{ s} - i{\rm \widetilde {\bf T}}_{ s} :{\rm
\widetilde{\bf \sigma }}
\end{equation}
\begin{equation}
\label{eq45}  [ \omega  - \alpha ({\rm {\bf k}} \cdot {\rm {\bf
v}}_0 ) - \alpha c({\rm {\bf k}} \cdot {\rm  {\bf{\bm \beta
}}})]{\rm {\bf B}}_{ s} = \alpha ({\rm {\bf k}} \cdot {\rm {\bf
v}}_{ { s}} ){\rm {\bf B}}_0 -\alpha ({\rm {\bf k}} \cdot {\rm
{\bf B}}_0 ){\rm {\bf v}}_{ { s}},
\end{equation}
\begin{equation}
\label{eq46}{\rm {\bf k}}\cdot {\rm {\bf B}}_{ { s}} = 0,
\end{equation}
where $\bf k$ is the wave vector and $\omega$ is the the oscillation
frequency. When Eqs.~(37) and (38) are substituted into
Eqs.~(39)-(41) and ${\bf v}_0 {\parallel} {\bf B}_0 {\parallel} {\bm
\beta}$ is considered, Eqs.~(39)-(41) can be converted to be,
\begin{eqnarray}
\label{eq47} {\rm \widetilde{\bf T}}_{ s} =(\frac{\Gamma p_0}{\rho_0}
\rho_{ s} + {\bf B}_0 \cdot {\bf B}_{s})\rm\widetilde{\bf I}+
\frac{\psi_{ s}}{\emph{c}^2}\gamma^2{\bf V}_0 {\bf V}_0 +
\frac{\psi_0}{\emph{c}^2}\gamma^2({\bf V}_0 {{\bf V}_{\mit s}}
+{{\bf V}_{\mit s}} {\bf V}_0)
 -({\bf B}_0{{\bf B}_{\mit s}}+{{\bf B}_{\mit s}}{\bf B}_0 ),
\end{eqnarray}
\begin{equation}
\label{eq48} {\rm {\bf P}_{ s}}= \frac{\gamma^2}{c^2}(\psi_0 {\bf v}_{ s}
+ \psi_{ s} {\bf v}_0) - \frac{1}{c^2}({\bf v}_0 \cdot {\bf B}_0){\bf
B}_{ s} + \frac{1}{c^2}({\bf B}_{ s} \cdot {\bf B}_0){\bf v}_0 -
\frac{1}{c^2}({\bf v}_{ s} \cdot {\bf B}_0){\bf B}_0 +
\frac{B_0^2}{c^2}{\bf v}_{ s},
\end{equation}
\begin{equation}
\label{eq49} \varepsilon=\psi_{ s} \gamma^2- \frac{\Gamma
p_0}{\rho_0}\rho_{s} -\gamma \rho_{ s} c^2+
{\bf B}_0 \cdot {\bf B}_{ s}.
\end{equation}
If $\omega = \alpha {\rm {\bf k}} \cdot {\rm {\bf v}}_0 + \alpha
c{\rm {\bf k}} \cdot {\rm {\bm \beta }}$, we get a unphysical
solution ($\omega = 0$) from Eqs.~(42)-(49), because the conditions
${\bf k} \cdot {\bf v}_s=0$ and ${\bf k} \cdot {\bf v}_0=0$ are
derived from Eqs.~(42)-(44). Now we consider $\omega \neq \alpha
{\rm {\bf k}} \cdot {\rm {\bf v}}_0 + \alpha c{\rm {\bf k}} \cdot
{\rm {\bm \beta }}$ and discuss all the three types of MHD waves:
the Alf\'ven wave, the ion-acoustic wave, and the magnetosonic wave.

\subsubsection{The Alf\'ven wave}

The transportation direction of Alf\'ven waves is along the magnetic
field line, so ${\bf k}{\overrightarrow
\parallel}{\bf v}_0{
\parallel} {\bf B}_0 {
\parallel} {\bm \beta}$,
where $\parallel$ denotes parallel and ${\overrightarrow\parallel}$
the same direction.
%We consider the couple with the photons in the
%energy release region close to the ISCO so ${\bf k}$ is supposed
%with the same direction of the velocity, although it doesn't affect
%our mathematical results if they are either in the same direction or
%in the contrary direction. The second is that
Since the Alf\'ven wave is a transverse wave, i.e. ${\bf k} \perp
{\bf B}_{s}$ , ${\bf k}\perp{\bf v}_{s}$ or the $\varphi$ component
of the perturbed velocity, $v_{ s,\varphi}=0$.

From Eqs.~(15)-(17), (42), (47), and (48) and the above results we
obtain \[{\bf f}_{{\rm curv},s}= (0,0,0),\] \[{\bf P}_{ s} \cdot
{\widetilde{\bf \sigma}}= \frac{\gamma^2}{c^2}(\psi_0 v_{
s,\varphi}+\psi_{ s} v_0) {\bf N} =0,\] and \[{\widetilde {\bf
T}}:{\widetilde {\bf \sigma}}=
\sigma_{31}[\frac{\gamma^2}{c^2}\psi_0 v_0+\frac{\alpha k
B_0^2}{\omega-\alpha({\bf k}\cdot {\bf v}_0)-\alpha c ({\bf k}\cdot
{\bm \beta})}]v_{s,\rm r},\] where $v_{s,\rm r}$ denotes the $r$
component of the perturbed velocity. Substituting the above
equations and Eqs.~(42), (45)-(49) into Eqs.~(43) and (44) we can
get $v_{ s,\rm r}=0$ and the dispersion equation,
\begin{equation}
\label{eq50} [\omega - \alpha c({\rm {\bf k}}\cdot {\rm {\bm \beta
}})]^2(\gamma ^2 \psi_0 + B_0^2 ) - (2\alpha \gamma ^2 \psi_0
v_0 k)[\omega - \alpha c({\rm {\bf k}}\cdot {\rm {\bm \beta }})] +
\alpha ^2\gamma ^2\psi _0 (k^2v_0^2 ) - \alpha ^2c^2k^2B_0^2 = 0.
\end{equation}
The frequencies of the Alf\'ven waves are solved as,
\begin{equation}
\label{eq51} \omega = k\alpha [\beta_3 c + \frac{\gamma ^2\psi_0 v_0
\pm B_0\sqrt {B_0^2c^2 + (c^2 - v_0^2)\gamma ^2\psi_0 } }{B_0^2 +
\gamma ^2\psi_0 }].
\end{equation}
The group velocities of the Alf\'ven waves are in the same form of
the phase velocities of the Alf\'ven waves and are,
\begin{equation}
\label{eq52} \frac{\omega}{k}=\frac{d\omega}{dk}  = \alpha [\beta_3
c + \frac{\gamma ^2\psi_0 v_0 \pm B_0\sqrt {B_0^2c^2 + (c^2 -
v_0^2)\gamma ^2\psi_0 } }{B_0^2 + \gamma ^2\psi_0 }].
\end{equation}
These velocities can be simplified in the special relativity, i.e.
$r \rightarrow\infty $, $\beta_3\rightarrow 0$ $\&$ $\alpha
\rightarrow 1$, as
\begin{equation}
\label{eq53} v_{\rm A}  = \frac{\gamma^2 v_0\pm \sqrt {\frac{B_0^2}{\psi
_0 } + (\frac{B_0^2}{\psi_0 })^2} c}{\frac{B_0^2}{\psi_0 } +
\gamma^2} = \frac{v_0\pm \eta\sqrt {\frac{1}{\gamma^2 } + \eta^2}
c}{ \eta^2 + 1},
\end{equation}
where $\gamma^2 \eta^2 = B_0^2/\psi_0 $. Equation (53) is in the
same form as the expression of De Villiers \& Hawley (2003).

\subsubsection{The ion-acoustic wave and the
magnetosonic wave}

The ion-acoustic wave is a longitudinal wave without electromagnetic
polarization so ${\bf k} {
\parallel} {\bf v}_{ s}$ and $B_{ s} = 0$. According to Eq.~(45) we have
${\bf B}_0 {\parallel} {\bf V}_{ s}$ owing to $B_{ s} = 0$. This
leads to the conclusion that ${\bf k} {
\parallel} {\bf v}_{s} {
\parallel} {\bf B}_0 {
\parallel} {\bf v}_0 {
\parallel} {\bm \beta}$, and that no ion-acoustic wave solution fits
Eqs.~(42)-(49).

The magnetosonic wave is another type of longitudinal wave with
transverse electromagnetic polarization, i.e., ${\bf k} \parallel
{\bf v}_{s}$, ${\bf k} \perp {\bf B}_{ s}$. The wave vector $\bf k$
isn't parallel with ${\bf B}_0$ under the condition ${\bf B}_0 {
\parallel} {\bf v}_0 {
\parallel} {\bm \beta}$, otherwise the wave is degenerated into the
ion-acoustic wave and does not exist. Now we discuss the
magnetosonic wave under two different conditions respectively.

Firstly when ${\bf k} \perp {\bf v}_0$ we get ${\bf k} \parallel
{\bf v}_s$, ${\bf k} \perp {\bf B}_{ s}$, ${\bf B}_0 {
\parallel} {\bf v}_0 {
\parallel} {\bm \beta} \perp {\bf k}$, and Eqs.~(42)-(49) can
be simplified to be,
\begin{equation}
\label{eq54} (\alpha \gamma^2 \psi_0 + \alpha \Gamma p_0 - \alpha
\gamma^2 \Gamma p_0 +\alpha B_0^2)({{\bf k} \cdot {\bf v}_{\rm s}}) = i
{\widetilde {\bf T}_{ s}} : {\widetilde {\bf \sigma}},
\end{equation}
\begin{equation}
\label{eq55} \omega^2 (\gamma^2 \psi_0 + B_0^2) {\bf v}_{s} +
\gamma^2 \alpha \Gamma p_0 ({\bf k}\cdot {\bf v}_{ s}) \omega {\bf v}_0
-\alpha^2 c^2 k^2 (2B_0^2 + \gamma^2 \rho_0 c^2 + \gamma^2
\frac{\Gamma^2}{\Gamma-1}p_0) {\bf v}_{ s} = i c^2 \omega (\alpha {\bf
f}_{{\rm curv},s} - {\bf P}_{s} \cdot {\widetilde {\bf \sigma}}),
\end{equation}
\begin{equation}
\label{eq56} {\widetilde {\bf T}_{ s}} : {\widetilde {\bf \sigma}} =
\frac{\gamma^2}{c^2} \frac{h_3}{h_1} \frac{\partial
\omega_3}{\partial x^1}\psi_0 v_0 v_{s,\rm r},
\end{equation}
\begin{equation}
\label{eq57} {\bf P}_{ s} \cdot {\widetilde {\bf \sigma}} =
\frac{\gamma^2}{c^2}  \alpha (\rho_0 c^2 + \frac{\Gamma^2}{
\Gamma-1}p_0) \frac{v_0}{\omega} ({\bf k}\cdot{\bf v}_{ s}) {\bf N},
\end{equation}
where
\begin{equation}
{\bf f}_{{\rm curv},s} = ( f_{s,\rm r}, 0,
G_{31}\frac{\gamma^2}{c^2} \psi_0 v_0 v_{s,\rm r}),
\end{equation}
with
\[
\label{eq58} f_{s,\rm r} = - G_{21} \alpha({\bf k}\cdot{\bf v}_{s})
\frac{\Gamma p_0 + B_0^2}{\omega} - G_{31} [\alpha \frac{\Gamma p_0
- B_0^2}{\omega} + \frac{\gamma^2}{c^2} \alpha (\rho_0 c^2 +
\frac{\Gamma^2}{ \Gamma-1}p_0) \frac{v_0^2}{\omega} ]({\bf
k}\cdot{\bf v}_{ s}).
\]
From Eqs.~(54)-(58) we can obtain an unstable  solution (an
increasing wave or an attenuation wave) if \[\frac{{\widetilde {\bf
T}_{s}} : {\widetilde {\bf \sigma}}}{\alpha \gamma^2 \psi_0 + \alpha
\Gamma p_0 - \alpha \gamma^2 \Gamma p_0 +\alpha B_0^2} = -\frac{c^2
f_{s,\varphi}}{\Gamma\gamma^2 p_0 v_0} = \frac{{\bf k}\cdot {\bf
v}_{s}}{i},\] or  no solution when ${\bf k} \perp {\bf v}_0$. The
possible unstable solution is
\begin{equation}
\label{eq59} \omega = \frac{A + \sqrt{A^2 + 4(\gamma^2\psi_0 +
B_0^2)(2B_0^2 + \gamma^2 \rho_0 c^2 +\frac{\gamma^2\Gamma^2
}{\Gamma-1}p_0)\alpha^2 c^2 k^2 }}{2\gamma^2\psi_0 + 2B_0^2},
\end{equation}
where \[A=\frac{c^2 f_{ s,\varphi}}{\Gamma\gamma^2 p_0 v_0
v_{ s}}\frac{c^2 \alpha f_{\rm s,r}-c^2 ({\bf P}_{\rm s} \cdot {\widetilde {\bf
\sigma}})_{\rm r}}{{\bf k}\cdot {\bf v}_{ s}},\] and $({\bf P}_{ s} \cdot
{\widetilde {\bf \sigma}})_{\rm r}$ represents the $r$ component of the
vector ${\bf P}_{ s} \cdot {\widetilde {\bf \sigma}}$.

Secondly when $\bf k$ is neither parallel with nor vertical to ${\bf
v}_0$ we  get the solution from Eq.~(42)-(49) as follows,
\begin{equation}
\label{eq60} \omega = \alpha c ({\bf k}\cdot{\bm \beta}) -
\frac{i\alpha ({\bf k}\cdot {\bf v}_0) ({{\widetilde {\bf T}_{
s}}}:{{\widetilde {\bf \sigma}}}) + 2\alpha^2 \gamma^2 \Gamma p_0
({\bf k}\cdot {\bf v}_{ s})({\bf k}\cdot {\bf v}_0)}{(\alpha
\gamma^2 \psi_0 + \alpha \Gamma p_0 - \alpha \gamma^2 \Gamma p_0
+\alpha B_0^2)({\bf k}\cdot {\bf v}_{ s})-\alpha ({\bf k}\cdot {\bf
B}_0)({\bf B}_0\cdot {\bf v}_{ s})-i {\widetilde {\bf T}_{ s}} :
{\widetilde {\bf \sigma}}},
\end{equation}
which is also unstable.

We summarize the results in \S2.1.1 and \S2.1.2: (1) there are two
stable Alf\'ven wave modes, (2) the ion-acoustic wave doesn't exist,
and (3) a few unstable modes of magnetoacoustic wave may emerge in
the GRMHD in the BH accretion disks.

\subsection{The relation between the magnetic energy density and the relativistic enthalpy density}

It is widely believed that the toroidal magnetic field in accretion
disks is generated by dynamo mechanism (Ruzmaikin et al. 1979; Tout
et al. 1992; Ruediger et al. 1995; Hawley et al. 2000; Moss et al.
2004), and that accretion is driven by the magnetic stress (e.g.
Matsumoto \& Tajima 1995; Brandenburg et al. 1995; Stone et al.
1996). Accordingly  the angular momentum conservation gives
(Torkelsson 1998)
\begin{equation}
\label{eq61}
\dot{M} r v_\varphi = 2 \pi rHr\frac{B'_\varphi B'_{\rm r} }{\mu _0 },
\end{equation}
where $H$ is the thickness of the accretion disk, $B'_\varphi$  and
$B'_{r}$ are the $\varphi$- and $r$-components of the magnetic field
$B'$, $\dot{M}(= 2\pi H r v_r \rho)$ is the accretion rate,
$v_\varphi$ and $v_{r}$ are the $\varphi$- and $r$-components of the
velocity of the accreting plasma, respectively. Suppose $B'_\varphi
= \gamma_{\rm dyn} B'_{r}$ and $v_{r} = l v_\varphi$, the above
equation can be simplified to be,
\begin{equation}
\label{eq62}{B'_\varphi}^2 = l v_\varphi ^2 \rho \mu _0 \gamma
_{\rm dyn}.
\end{equation}
The velocity $v_\varphi$ can be approximatively  expressed as the
velocity of the circular orbit relative to Bardeen observers, which
can be got from the equations (8.354-8.359) of Camenzind (2007). The
velocity of the plasm in the prograde orbit is,
\begin{equation}
\label{eq63}v_\varphi = [(i^2 + a_\ast ^2 + \textstyle{2 \over
i}a_\ast ^2 )\frac{i^2 + a_\ast ^2 - 2i}{i^3\sqrt i - 2i^2\sqrt i +
a_\ast ^2 i\sqrt i + a_\ast i^2 + a_\ast ^3 - 2a_\ast i} -
\textstyle{2 \over i}a_\ast ]\frac{c}{\sqrt {i^2 - 2i + a_\ast ^2 }
},
\end{equation}
and in the retrograde orbit,
\begin{equation}
\label{eq64}v_\varphi = [(i^2 + a_\ast ^2 + \textstyle{2 \over
i}a_\ast ^2 )\frac{4a^\ast \sqrt{i}-(i^2 + a_\ast ^2 - 2i)}{i^3\sqrt
i - 2i^2\sqrt i + a_\ast ^2 i\sqrt i - a_\ast i^2 - a_\ast ^3 +
2a_\ast i} - \textstyle{2 \over i}a_\ast ]\frac{c}{\sqrt {i^2 - 2i +
a_\ast ^2 } },
\end{equation}
where $a_* = |a|$, $i= r/r_{\rm g}$. The two velocities return to
the Keplerian velocity $c/\sqrt{i}$ when $i \rightarrow \infty$.

Since $\frac{\Gamma}{\Gamma-1}p_0 \ll \rho_0 c^2$ in the accretion
disks, we can express $\frac{B_\varphi^{'2}}{\mu_0}=\gamma^2\eta^2
(\rho_0 c^2 + \frac{\Gamma p_0}{\Gamma - 1})$ as
$\frac{B_\varphi^{'2}}{\mu_0} = B_\varphi^2 \simeq \gamma^2\eta^2
\rho_0 c^2$ and $\gamma^2\eta^2 = \gamma_{\rm dyn}l
v_\varphi^2/{c^2} = l' v_\varphi^2/{c^2}$ from Eq.~(62), where $ l'
=  \gamma_{\rm dyn}l = \frac{B'_\varphi}{B'_{\rm r}}\frac{v_{\rm
r}}{v_\varphi}$. We then get the relation between the magnetic
energy density and the relativistic enthalpy density,
$\frac{B_{\varphi}^{'2}}{2\mu_0} \simeq \frac{l \gamma_{\rm
dyn}}{2}\frac{v_\varphi^2}{c^2}\psi_0$, and express the frequencies
of the Alf\'ven waves, i.e., Eq.~(51), as,
\begin{equation}
\label{eq65} \omega = k\alpha [\beta_3 c + \frac{\gamma^2 \pm
\sqrt{l^{'}+{l^{'}}^2 \frac{v_\varphi^2}{c^2}}}{\gamma^2 +
l^{'}\frac{v_\varphi^2}{c^2} }v_\varphi].
\end{equation}

When we consider the characteristic wavelength $\lambda \sim r$ ,
i.e. the wave number $k \sim 2\pi/r$, Eq.~(65) is reduced to be
\begin{equation}
\label{eq66} \omega \simeq \frac{2\pi}{r} c \xi = \frac{2\pi}{i r_g}
c \xi
\end{equation}
where $\xi = \alpha [\beta_3  + \frac{\gamma^2 \pm
\sqrt{l^{'}+{l^{'}}^2 v_\varphi^2/c^2}}{\gamma^2 \pm
l^{'}v_\varphi^2/c^2 }\frac{v_\varphi}{c}]$.

\subsection{Estimate of the parameter $l'$}

Begelman and Pringle (2007) have investigated the structure of
accretion disks with strong toroidal magnetic fields, and found that
the  thickness of the disks is higher than standard thin disks, but
in line with observations (Robinson et al. 1999; Shafter \& Misselt
2006),
\begin{equation}
\label{eq67}\frac{H}{r} = 0.48 \alpha_*^{-1/17}{r_{10}^{9/68}}
\dot{M}_{18}^{3/34}(\frac{M}{M_\odot})^{-15/68},
\end{equation}
where $\alpha_*$ is the viscosity prescription,
$\dot{M}_{18}=\dot{M}/10^{18}$ gs$^{-1}$, and $r_{10}=r/10^{10}$ cm.
The corresponding radial velocity is,
\begin{equation}
\label{eq68}v_{\rm r} \simeq \frac{3}{2} \alpha_* (\frac{H}{r})^2 v_{\rm k},
\end{equation}
where $v_{\rm k}$ is the Keplerian velocity. Here we adopt that
$v_\varphi\simeq v_{\rm k}$ to estimate the value of $l'$. From Eqs.
(67) and (68), the parameter $l$ is,
\begin{equation}
\label{eq69} l = \frac{v_{\rm r}}{v_\varphi} \simeq 0.3456 \alpha_*
^{15/17}{r_{10}^{9/34}}
\dot{M}_{18}^{3/17}(\frac{M}{M_\odot})^{-15/34}.
\end{equation}
Vishniac et al.(1990) have discussed the dynamo action by internal
waves in accretion disks and suggested $B_{\rm r}/B_{\varphi} \sim
\alpha_* $, so we have
\begin{equation}
\label{eq70} l' = l \gamma_{\rm dyn} \simeq 0.3456 \alpha_*
^{-2/17}{r_{10}^{9/34}}\dot{M}_{18}^{3/17}(\frac{M}{M_\odot})^{-15/34}.
\end{equation}
The relatively small values of the power indices in Eq.~(70) indicate that
$l'$ is not sensitively dependent on $\alpha_*$, $r$, $M$ and $\dot{M}$.

HFQPOs in BH-LMXBs were generally observed in the steep
power-law (SPL) state, %or%
i.e. very high state (VHS), and it is very likely that the QPOs are
associated with the region that produces hard X-ray emission. The
accretion model for the VHS is a long-debated subject. It might
contain an inner ADAF surrounded by an outer thin disk (e.g. Yuan
2001). By analyzing the observational data of GRO J1655$-$40 and XTE
J1550$-$564, McClintock \& Remillard (2006) showed that the disks
with blackbody radiation appear to truncate at a radius ($\sim 100
r_{\rm g}$) in the low/hard state, and that the truncated radius
decreases when the power-law component becomes stronger and steeper.
Theoretical investigations (Abramowicz et al. 1995; Honma 1996; Esin
et al. 1997; Liu et al. 1999; Manmoto et al. 2000; R\'o\`za\'Nska et
al. 2000; Narayan et al. 2008) also suggest that the transition
radius ranges from $\sim 100 r_{\rm g}$ to $10000 r_{\rm g}$.
Accordingly the transition radius in VHS is likely to be smaller
than in the low/hard state, and for illustration here we adopt its
value to be $50r_{\rm g}$. The X-ray luminosity in the VHS is often
more than $0.2L_{\rm Edd}$ where $L_{\rm Edd}$ is the Eddington
luminosity (McClintock \& Remillard 2006), so we adopt a typical
accretion rate $10^{18}$ gs$^{-1}$. We also take $\alpha_*\sim 0.1$
(King et al. 2007). From the above values we can estimate the
parameter $l'$ as,
\begin{equation}
\label{eq71} l'\sim 0.048 (\frac{\alpha_*}{0.1})^{-2/17}
(\frac{r}{50r_{\rm g}})^{9/34}(\frac{\dot{M}}{10^{18}\,{\rm
gs}^{-1}})^{3/17} (\frac{M}{7M_{\odot}})^{-15/34}.
\end{equation}

Combing Eqs.~(3), (4), (66), and (71) we get the two frequencies of
the HFQPOs as,
\begin{equation}
\label{eq72} \omega = 1.2756 \times {10^6} \frac{\xi}{i}
(\frac{M}{M_\odot})^{-1}
\end{equation}
where \[\xi = \sqrt{\frac{i^4-2i^3+a^2i^2}{i^4+a^2i^2+2a^2i}}
[\frac{2a}{i\sqrt{i^2-2i+a^2}}  + \frac{\gamma^2 \pm
\sqrt{0.04765+0.0022705 v_\varphi^2/c^2}}{\gamma^2 + 0.04765
v_\varphi^2/c^2 }\frac{v_\varphi}{c}].\] Our calculations show that,
for BH-LMXBs with measured HFQPOs the ratio of the two frequencies
is generally around 1.5 (see Table 1).

\section{ Results and discussion}

Equations (63), (64), and (72) indicate the existence of Alf\'ven
waves with two frequencies in the accretion disk. The ratio of the
upper and lower frequencies is close to $3:2$, suggesting that these
waves may account for the HFQPO pairs. Given the mass and spin of a
BH, one can determine the radius where the QPOs are produced. In
Table 1 we present the inferred radius by comparing Eq.~(72) with
either the upper or the lower centroid QPO frequency of several
BH-LMXBs. We adopt the averaged masses for GRO J1655$-$40
($6.0-6.6M_\odot$, McClintock \& Remillard 2006), GRS 1915+105
($10-18M_\odot$, Greiner et al. 2001), and XTE J1550$-$564
($8.4-10.8 M_\odot$,  McClintock \& Remillard 2006) and averaged
dimensionless spins of GRO J1655$-$40 ($0.65-0.75$, Shafee et al.
2006) and GRS 1915$+$105 ($0.98-1.0$, McClintock \& Shafee et al.
2006). Because the spin of XTE J1550$-$564 hasn't been measured, we
take it to be 1, 0.5, and 0. It is interesting to see that in most
cases the radii are $\sim 70 r_{\rm g}$, consistent with the
transition radii between an ADAF and a thin disk discussed above.
The frequencies of the HFQPOs mainly depends on the transition
radius ($r_{\rm tr}$)  and the ratio of $l'=(v_{\rm
r}/v_{\varphi})/(B_{\rm r}^{'}/B_{\varphi}^{'})$. According to the
discussion in \S2.3 the parameter $l'$ changes little with the
$\alpha_*$,  $r$, $M$, and $\dot{M}$, this may explain why the HFQPO
frequencies are relatively stable during the VHS.

Considering the similarities in BH and NS accretion disks, it is
interesting to ask why the 1.5 frequency ratio is not evident in the
HFQPOs in NS-LMXBs. The reasons may lie in the differences in the
configuration of the magnetic fields and the structure of the
accretion disks  in BH- and NS-LMXBs. It is well known that NSs
generally hold dipolar magnetic fields, and the toroidal field
component, induced by difference in the angular velocity between the
disk and the NS magnetosphere, can never be stronger than the
poloidal component, otherwise the field configuration becomes
unstable leading to inflation of the field lines (e.g. Aly 1985). On
the other hand, both theories and simulations (Ruzmaikin et al.,
1979; Tout et al., 1992; Ruediger et al., 1995; Hawley et al., 2000;
Moss et al, 2004; Hirose et al., 2004) show that in BH accretion
disks the toroidal magnetic field can be much stronger than the
poloidal one. Additionally, recent observations suggest that the
inner disk around an NS may not be an ADAF even in low state, as the
thermal emission from the surface of the NS would tend to cool an
ADAF, making such a flow more difficult than in BH systems (Cackett
et al., 2009).

Finally, a correlation between the lower HFQPO frequencies and the
BH masses has been suggested, i.e., $\nu_{\rm l}\propto M^{-1}$
(McClintock \& Remillard 2006). This correlation can be naturally
reproduced in our model. In Fig.~1 we plot the predicted relation
between $\nu_{\rm l}$ and $m$ when $r_{\rm tr}$ changes from $66 r_{
g}$ to $76 r_{ g}$, which fit reasonably with the measured data.

\acknowledgements

We are grateful to an anonymous referee for helpful comments. This
work was supported by the Natural Science Foundation of China (under
grant number 10873008) and the National Basic Research Program of
China (973 Program 2009CB824800).

\clearpage
\begin{table*}[h]
 \centering
  \caption{This table presents the masses ($m$ in solar units), dimensionless spins $a$ for several
  BHs, and the inferred radii ($r_{\rm tr, u}$ and $r_{\rm tr, l}$), the
predicted frequency ratios ($\nu_{\rm u}/\nu_{\rm l}$) in the
corresponding sources based on the measured upper and lower QPO
frequencies ($\nu_{\rm u}$ and $\nu_{\rm l}$ in Hz) respectively. }
\label{t:1}
  \begin{tabular}{@{}lcccccccc@{}}
   \hline
  \hline
  {sources} & $m $ & {$a$}
&{$\nu_{\rm u}$} & {$r_{\rm tr,u}$} & $\nu_{\rm u}/\nu_{\rm l}$&
{$\nu_{\rm l}$} &{$r_{\rm tr,l} $} & $\nu_{\rm u}/\nu_{\rm l}$\\
 \hline
GRO J1655-40 & 6.3 &  0.7 & 450 & 66.7528  &1.5504& 300 & 65.3002 & 1.5501\\

GRS 1915+105 & 14 & 0.99 & 168 & 75.6114 &1.5514& 113 & 73.4984  & 1.5510\\

GRS 1915+105 & 14 & 0.99 & 67 & 139.7920  &1.5570& 41 & 144.368  & 1.5572\\

XTE J1550-564 & 9.6 & 1 & 276 & 69.8141  &1.5503& 184 & 68.2975   & 1.5500\\

XTE J1550-564 & 9.6 & 0.5 & 276 & 69.8639  &1.5515 & 184 & 68.3125 & 1.5512\\

XTE J1550-564 & 9.6 & 0 & 276 & 69.9159  &1.5526 & 184 &68.3302  & 1.5524\\
\hline
\end{tabular}
\end{table*}

\begin{center}
\begin{figure}[h]
\includegraphics[width=0.7\textheight]{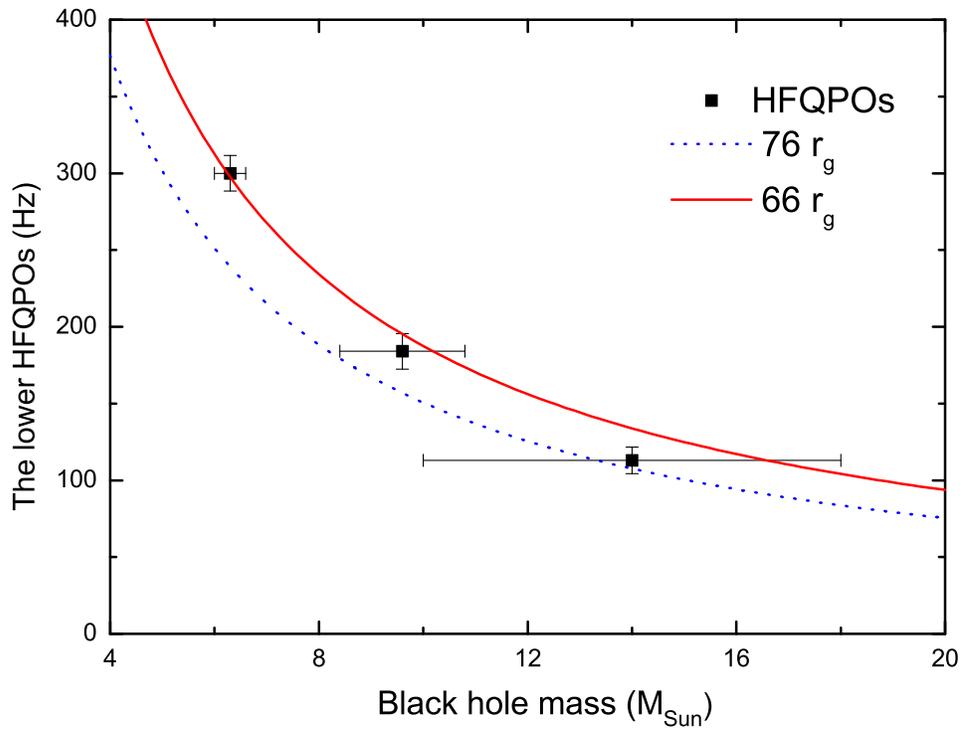}
  \caption{ The relation between the lower HFQPO frequencies and BH masses
  for GRO J1655-40, XTE J1550-564, and GRS 1915+105.} \label{fig1}
\end{figure}
\end{center}

\end{document}